\documentclass[aps,prb,twocolumn,superscriptaddress,showpacs]{revtex4}

\usepackage{amsmath}
\usepackage{bm}
\usepackage{graphicx}

\begin{document}

\title{Ghost Fano resonance in a double quantum dot molecule attached to leads}

\author{M. L. Ladr\'on de Guevara}
\affiliation{Facultad de F\'{\i }sica, Pontificia Universidad
Cat\'{o}lica de Chile, Casilla 306, Santiago 22, Chile}
\affiliation{Departamento de F\'{\i }sica, Universidad
Cat\'{o}lica del Norte, Casilla 1280, Antofagasta, Chile}

\author{F. Claro}
\affiliation{Facultad de F\'{\i }sica, Pontificia Universidad
Cat\'{o}lica de Chile, Casilla 306, Santiago 22, Chile}

\author{Pedro A. Orellana}
\affiliation{Departamento de F\'{\i }sica, Universidad
Cat\'{o}lica del Norte, Casilla 1280, Antofagasta, Chile}

\begin{abstract}
We study the electronic transport through a double quantum dot
molecule attached to leads, and examine the transition from a
configuration in series to a symmetrical parallel geometry. We
find that a progressive reduction of the tunneling through the
antibonding state takes place as a result of the destructive
quantum interference between the different pathways through the
molecule. The Fano resonance narrows down, dissapearing entirely
when the configuration is totally symmetric, so that only the
bonding state participates of the transmission. In this limit the
antibonding state becomes completely localized.
\end{abstract}

\pacs{73.21.La, 73.23.-b, 73.63.Kv}

\maketitle

\section{Introduction}

The tunneling of electrons through quantum dot structures has been
the subject of active research during the last years. For the
confinement of electron in all three dimensions, quantum dots are
characterized by the discreteness of the energy levels, and for
this reason are often called ``artificial atoms''. \cite{qd1,qd2}
One of the main features of transport through quantum dots is that
the coherence of electrons is greatly preserved, as manifested in
phenomena such as the Aharonov-Bohm oscillations in multiply
connected geometries, \cite{ABoscill} the Kondo effect in dots
strongly coupled to electron reservoirs, \cite{kondo} and
Fano-type line shapes in transport through multiple channels.
\cite{fano}

Two or more quantum dots can be coupled to form an ``artificial
molecule'', in which electrons are shared by the different sites.
The formation of bonding and antibonding states in such molecules
has been studied by means of transport
\cite{blick2,jeong,holleitner1,holleitner2} as well as
spectroscopy experiments. \cite{schedelbeck,oosterkamp}
Theoretical work on electron transport through serial quantum dot
configurations is contained in Refs. \cite{tmols}, and quantum
interference effects have been explored in parallel and
``T-shaped'' geometries. \cite{Tdqd1,kang,boese} Particular
interest in quantum dot molecules lies in their potential
application in quantum computing devices. In this context, diverse
proposals have been made, where the quantum bits are built with
electron spin states \cite{loss1} or with the coherent mode
generated by discrete states in an artificial molecule.
\cite{blick3}

In this work we study the electronic transport through a double
quantum dot molecule attached to leads, in a transition from a
connection in series to a completely symmetrical parallel
configuration. We examine the linear conductance at zero
temperature and obtain the associated densities of states of each
of the dots. We find that the conductance spectrum is composed of
a Lorentzian centered at the bonding energy and a Fano line shape
at the energy of the antibonding state. The latter arises due to
the presence of a bound state of the molecule, immersed in the
band continuum. A progressive line narrowing of the Fano peak is
observed as the system transits from the series to the symmetrical
parallel configuration. For the perfectly symmetrical geometry,
the antibonding state is totally uncoupled of the leads, the
bonding state becoming the only one that participates in the
transmission. In this extreme case, the antibonding state is
localized, with zero localization length. Although exponentially
localized states generally exist outside the band continuum, this
state is in the conduction band and acts like a ghost of the Fano
resonance.

\section{Model}

\begin{figure}[h]
\par
\begin{center}
\includegraphics[scale=0.38,angle=-90]{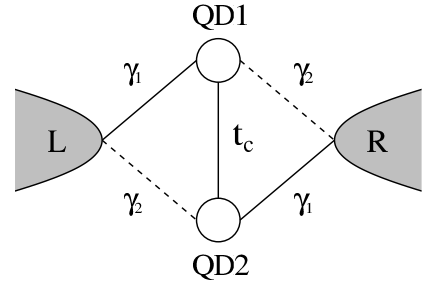}
\end{center}
\caption[Fig1]{Scheme of a quantum dot molecule coupled to left (L) and right
(R) leads.}
\label{Fig1}
\end{figure}

We consider two neighboring quantum dots forming a molecule,
coupled to left and right leads as shown in Fig. \ref{Fig1}. Only
one energy level in each dot is assumed relevant and the interdot
and intradot electron-electron interactions are neglected.
\noindent The system can be modelled by a non-interacting two-impurity
Anderson Hamiltonian, which can be written as
\begin{equation}
H=H_{m}+H_{l}+H_{I},  \label{eq-1}
\end{equation}
where $H_{m}$ describes the dynamics of the isolate molecule,
\begin{equation}
H_{m}=\sum_{i=1}^{2}\varepsilon _{i}d_{i}^{\dag }d_{i}-t_{c}(d_{2}^{\dag
}d_{1}+d_{1}^{\dag }d_{2}).  \label{eq-2}
\end{equation}
\noindent Here $\varepsilon _{i}$ is the energy of dot $i$, $d_{i}$ $%
(d_{i}^{\dagger })$ annihilates (creates) an electron in dot $i$ and $t_{c}$
is the interdot tunneling coupling. $H_{l}$ is the Hamiltonian for the
noninteracting electrons in the left and right leads
\begin{equation}
H_{l}=\sum_{k_{\alpha }\in \{L,R\}}\varepsilon _{k_{\alpha }}c_{k_{\alpha
}}^{\dag }c_{k_{\alpha }},  \label{eq-3}
\end{equation}
where $c_{k_{\alpha }}$ $(c_{k_{\alpha }}^{\dag })$ is the annihilation
(creation) operator of an electron of quantum number $k_{\alpha }$ and
energy $\varepsilon _{k_{\alpha }}$ in the contact $\alpha $. Finally, $%
H_{I} $ accounts for the tunneling between dots and leads,
\begin{eqnarray}
H_{I} &=&\sum_{k_{\alpha }\in \{L,R\}}V_{1k_{\alpha }}d_{1}^{\dag
}c_{k_{\alpha }}+\mbox{ h. c.}  \nonumber \\
&&+\sum_{k_{\alpha }\in \{L,R\}}V_{2k_{\alpha }}d_{2}^{\dag }c_{k_{\alpha }}+%
\mbox{ h. c.},  \label{eq-4}
\end{eqnarray}
with $V_{ik_{\alpha }}$ the tunneling matrix element. The linear
conductance $G$ is related to the transmission $T(\varepsilon )$
of an electron of energy $\varepsilon $ by the Landauer formula at
zero temperature \cite{datta},
\begin{equation}
G=\frac{2e^{2}}{h}T(\varepsilon _{F}).  \label{eq-5}
\end{equation}
To obtain explicitly $G$, we use the equation of motion approach
for the Green's function \cite{datta}. The retarded Green's
function is defined by
\begin{equation}
G_{ij}^{r}(t)=-i\theta (t)\langle \{d_{i}(t),d_{j}^{\dagger }(0)\}\rangle
,\quad i,j=1,2,  \label{eq-6}
\end{equation}
where $\theta (t)$ is the step function. In the absence of interaction, the
total transmission $T(\varepsilon )$ can be expressed as
\begin{equation}
T(\varepsilon )=tr\{\mathbf{G}^{a}(\varepsilon )\mathbf{\Gamma }^{R}\mathbf{G%
}^{r}(\varepsilon )\mathbf{\Gamma }^{L}\},  \label{eq-7}
\end{equation}
where $\mathbf{G}^{r(a9)}(\varepsilon )$ is the Fourier transform of the
retarded (advanced) Green's function of the molecule, and the matrix $%
\mathbf{\Gamma }^{L(R)}$ describes the tunneling coupling of the two quantum
dots to the left (right) lead, with
\begin{equation}
\Gamma _{ij}^{L(R)}=2\pi \sum_{k_{L(R)}}V_{ik_{L(R)}}V_{jk_{L(R)}}^{*}\delta
(\varepsilon -\varepsilon _{k_{L(R)}}),\quad i,j=1,2.  \label{eq-8}
\end{equation}
The equation of motion method uses the Heisenberg equations for the Fermi
operators of the molecule, which are inserted in the time derivatives of the
Green's functions of Eq. (\ref{eq-6}). This leads to first order
differential equations for the $G_{ij}^{r}$'s, containing Green's functions
for different dot operators, as well as others involving dot and lead
operators. Through the same procedure, equations for these new Green's
functions are obtained, until having a closed set. A Fourier transform of
such a set of equation takes it to an algebraic linear system for the $%
G_{ij}^{r}(\varepsilon )$'s, the solution of which leads to the following
expression for $\mathbf{G}^{r}$
\begin{equation}
\mathbf{G}^{r}(\varepsilon )=\frac{1}{\Omega }\left(
\begin{array}{ll}
\varepsilon -\varepsilon _{2}+i\frac{\Gamma _{22}}{2} & \;\;\;\;-t_{c}+i%
\frac{\Gamma _{21}}{2} \\
\;\;\;\;-t_{c}+i\frac{\Gamma _{12}}{2} & \varepsilon -\varepsilon _{1}+i%
\frac{\Gamma _{11}}{2}
\end{array}
\right) ,  \label{eq-9}
\end{equation}
where
\begin{equation}
\Omega =(\varepsilon -\varepsilon _{1}+i\frac{\Gamma _{11}}{2})(\varepsilon
-\varepsilon _{2}+i\frac{\Gamma _{22}}{2})-(-t_{c}+i\frac{\Gamma _{12}}{2}%
)(-t_{c}+i\frac{\Gamma _{21}}{2}),  \label{eq-10}
\end{equation}
with $\Gamma _{ij}=\Gamma _{ij}^{L}+\Gamma _{ij}^{R}$. A useful
quantity which provides insight into the electronic distribution
is the local density of states in dot $i$. It can be written in
terms of the diagonal matrix elements of the retarded Green's
function,
\begin{equation}
\rho _{i}(\varepsilon )=-\frac{1}{\pi }{\mbox Im}G_{ii}^{r}(\varepsilon
),\quad i=1,2.  \label{eq-11}
\end{equation}

We are interested in the particular situation described in Fig.
\ref{Fig1}, in which $\Gamma _{11}^{R}=\Gamma _{22}^{L}\equiv
\gamma _{1}$ and $\Gamma _{11}^{L}=\Gamma _{22}^{R}\equiv \gamma
_{2}$, so that $\gamma _{2}=0$ represents a connection in series,
and $\gamma _{2}=\gamma _{1}$ a symmetrical configuration in
parallel. For this case, the non diagonal matrix elements of the
matrices $\mathbf{\Gamma }^{L,R}$ obey $\Gamma
_{21}^{L}=\Gamma _{12}^{L}=\Gamma _{21}^{R}=\Gamma _{12}^{R}\equiv \sqrt{%
\gamma _{1}\gamma _{2}}$.

\section{Conductance and local density of states}

Introducing Eqs. (\ref{eq-9})-(\ref{eq-10}) in Eq. (\ref{eq-7}) and then in (%
\ref{eq-5}), and using the above values of the $\mathbf{\Gamma
}^{L,R}$ matrix elements, we obtain the following expression for
the conductance
\begin{equation}
G(\varepsilon )=\frac{2e^{2}}{h}\frac{4}{C_{1}}\left[ t_{c}\bar{\gamma}%
-\gamma _{12}(\varepsilon -\bar{\varepsilon})\right] ^{2},  \label{eq-12}
\end{equation}
\noindent where
\begin{eqnarray}
C_{1} &=&\left[ (\varepsilon -\bar{\varepsilon})^{2}-(\Delta \varepsilon
/2)^{2}-t_{c}^{2}-\frac{(\Delta \gamma )^{2}}{4}\right] ^{2}  \nonumber \\
&+&4\left[ \bar{\gamma}(\varepsilon -\bar{\varepsilon})-t_{c}\gamma
_{12}\right] ^{2},  \label{eq-13}
\end{eqnarray}
with $\bar{\varepsilon}=(\varepsilon _{1}+\varepsilon _{2})/2$, $\Delta
\varepsilon =\varepsilon _{1}-\varepsilon _{2}$, $\bar{\gamma}=(\gamma
_{1}+\gamma _{2})/2$, $\Delta {\gamma }=\gamma _{1}-\gamma _{2}$ and $\gamma
_{12}=\sqrt{\gamma _{1}\gamma _{2}}$. The densities of states at each of the
quantum dots are, in turn,
\begin{widetext}
\begin{equation}
\rho _{1}(\varepsilon )=\frac{1}{\pi C_{1}}\left\{ \bar{\gamma}[t_{c}^{2}+%
\frac{(\Delta \gamma )^{2}}{4}+(\varepsilon -\overline{\varepsilon }+\frac{%
\Delta \varepsilon }{2})^{2}]+4t_{c}\gamma _{12}(\varepsilon -\overline{%
\varepsilon }+\frac{\Delta \varepsilon }{2})\right\} ,  \label{eq-16}
\end{equation}
\noindent and
\begin{equation}
\rho _{2}(\varepsilon )=\frac{1}{\pi C_{1}}\left\{ \bar{\gamma}[t_{c}^{2}+%
\frac{(\Delta \gamma )^{2}}{4}+(\varepsilon -\overline{\varepsilon }-\frac{%
\Delta \varepsilon }{2})^{2}]+4t_{c}\gamma _{12}(\varepsilon -\overline{%
\varepsilon }-\frac{\Delta \varepsilon }{2})\right\} ,  \label{eq-17}
\end{equation}
\end{widetext}
with $C_{1}$ the same as Eq. (\ref{eq-13}).

First, let us consider the case with $\Delta \varepsilon =0$. It follows from
Eqs. (\ref{eq-12})-(\ref{eq-13}) that $G(\varepsilon )$ has two resonances,
corresponding to the bonding and antibonding states, at the energies
\begin{equation}
\varepsilon _{\pm }=\bar{\varepsilon}\pm \sqrt{t_{c}^{2}-\frac{(\Delta
\gamma )^{2}}{4}},  \label{eq-14}
\end{equation}
which in general differ from the eigenvalues for the isolate molecule, $%
\varepsilon _{\pm }^{0}=\bar{\varepsilon}\pm t_{c}$. The
resonances can be distinguished only when $|t_{c}|>|\Delta \gamma
|/2$, and the separation between them decreases as $|\Delta \gamma
|/2$ approaches $|t_{c}|$. Thus, the level attraction produced by
the connection to the leads becomes smaller as the coupling
strength $\gamma _{2}$ increases, up to vanishing when $\gamma
_{2}=\gamma _{1}$, where the bonding and antibonding states
coincide with those of the isolated molecule. A similar level
attraction is reported by Kubala \emph{et al.} in Ref.
\cite{ddotp1} in an Aharonov-Bohm ring with a quantum dot in each
of its arms.

We notice also that the conductance vanishes at the Fermi energy \begin{equation}
\varepsilon _{A}=\bar{\varepsilon}+t_{c}\bar{\gamma}/\gamma _{12},\quad
\label{antires}
\end{equation}
provided $\gamma _{2}\neq 0,\gamma _{1}$. This antiresonance-like
behavior, characterized by strictly zero transmission, is a
consequence of destructive quantum interference between the
different pathways through the dots, and does not exist for the
connection in series, where only one possible pathway exists.
Also, it can be shown that $G(\varepsilon )$ reaches the quantum
limit $2e^{2}/h$ at specific values of the energy provided
$|t_{c}|\geq |\Delta \gamma |/2$. For the molecule connected in
series perfect transmission requires that the interdot coupling
strength be at least $\gamma _{1}/2$, while in the parallel
configuration with a weak interdot coupling it behaves as an ideal
channel if the dot-lead coupling strengths are all of similar
magnitude. Without loss of generality we can set
$\bar{\varepsilon}=0$, convention we adopt in what follows.

\begin{figure}[h]
\centering
\includegraphics[scale=0.3,angle=0]{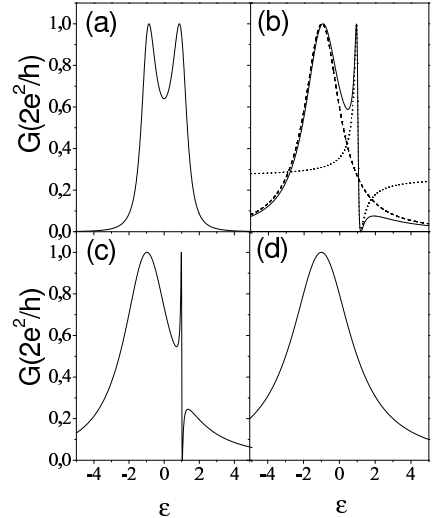}
\caption[Fig2]{Conductance as a function of the Fermi energy, for
$\Delta \varepsilon =0$, $t_{c}=\gamma _{1}$ and different values
of $\gamma _{2}$: (a) $\gamma _{2}=0$, (b) $\gamma _{2}=0.3\gamma
_{1}$, (c) $\gamma _{2}=0.6$ $\gamma _{1}$ and (d) $\gamma
_{2}=\gamma _{1}$. The dash and dotted lines in (b) correspond,
respectively, to the Breit-Wigner and Fano line shapes of Eqs.
(\ref{eq-19}) and (\ref{eq-20}). } \label{Fig2}
\end{figure}

In Fig. \ref{Fig2} we have plotted the conductance as a function
of the Fermi energy for $\Delta \varepsilon =0$, $t_{c}=\gamma
_{1}$, and different values of $\gamma _{2}$. For the
configuration in series (Fig. 2a) the bonding and antibonding
resonances are clearly resolved. For $\gamma _{2}\neq 0,\gamma
_{1}$, as in Figs. 2(b) and (c), the conductance exhibits the
antiresonance mentioned above, and there is a progressive increase
(reduction) of the width of the bonding (antibonding) resonance,
as $\gamma _{2}$ becomes larger ($\gamma _{2}<\gamma _{1}$). When
$\gamma _{2}=\gamma _{1\text{ }}$(Fig 2d) the antiresonance and
the peak associated to the antibonding state are no longer
present.

It can be shown that the conductance for $\Delta \varepsilon =0$
is composed by a Breit-Wigner and a Fano line shape centered at
the bonding and antibonding energies, respectively. Defining the
quantities $\Gamma _{-}$ and $\Gamma _{+}$ by
\begin{equation}
\Gamma _{\pm }=\bar{\gamma}\pm \gamma _{12},  \label{eq-18}
\end{equation}
in the limit $\Gamma _{+}\gg \Gamma _{-}$ and around $\varepsilon =-t_{c}$, $%
G(\varepsilon )$ can be approximated by
\begin{equation}
G(\varepsilon )\simeq \frac{2e^{2}}{h}\frac{\Gamma
_{+}^{2}}{\Gamma _{+}^{2}+(\varepsilon +t_{c})^{2}}, \label{eq-19}
\end{equation}
and around $\varepsilon = t_{c}$, by the Fano line shape of width
$\Gamma _{-}$
\begin{equation}
G(\varepsilon )\simeq \frac{2e^{2}}{h}\frac{\Gamma _{+}^{2}}{\Gamma
_{+}^{2}+4t_{c}^{2}}\frac{(Q+e_{-})^{2}}{%
1+e_{-}^{2}},  \label{eq-20}
\end{equation}
where $Q=-2t_{c}/\Gamma _{+}$ and $e_{-}=(\varepsilon -t_{c})/\Gamma _{-}$ ($\Gamma
_{-}\neq 0)$. The width of the bonding resonance $\Gamma _{+}$ ranges from $%
\gamma _{1}/2$ for the connection in series to $2\gamma _{1}$ in
the symmetrical case, while $\Gamma _{-}$ decreases from $\gamma
_{1}/2$ to become infinitely small when $\gamma _{2}$ approaches
$\gamma _{1}$. A similar line reduction in the conductance
spectrum is discussed in Ref. \cite {2channel} in a junction with
two resonant impurities.  This line narrowing (widening) in the
conductance can be interpreted as an increase (reduction) of the
lifetime $\tau =\hbar /\Gamma $ of the corresponding molecular
state, with $\Gamma$ the associated linewidth. The strong
narrowing of the Fano peak when $\gamma _{2}$ is close to $\gamma
_{1}$ is a signal of slow transitions between the antibonding
state and the leads. The lifetime of the antibonding state becomes
infinitely long when $\gamma _{2}$ approaches $\gamma_{1}.$
Typically, the $\gamma$'s are of the order of $meV$'s, so that the
lifetimes of quantum dots attached to leads in series are of the
order of $picoseconds$. However, if in the present configuration
$\gamma _{1}=10$ $meV$ and $\gamma _{2}=9.9$ $meV$, one obtains
$\Gamma _{-}=5.5\times 10^{-4}$ $meV$, then the lifetime of the
antibonding state is $\tau _{-}\simeq 33$ $ns$, that is, four
orders of magnitude larger. It is interesting to note that, as
follows from expressions (\ref{eq-12})-(\ref{eq-13}) and as
displayed by Fig. \ref{Fig2}(d), when $\gamma _{2}=\gamma _{1}$,
the Fano resonance entirely disappears and the conductance reduces
exactly to the Lorentzian
\begin{equation}
G(\varepsilon )=\frac{2e^{2}}{h}\frac{4\gamma
_{1}^{2}}{(\varepsilon +t_{c})^{2}+4\gamma _{1}^{2}},
\label{eq-15}
\end{equation}
expression with the form of the conductance of a single quantum
dot centered at the bonding energy, with an effective line
broadening $\gamma _{e}=2\gamma _{1}$. This would indicate that
only the bonding state contributes to the transmission through the
molecule.

\begin{figure}[h]
\centering
\includegraphics[scale=0.3,angle=0]{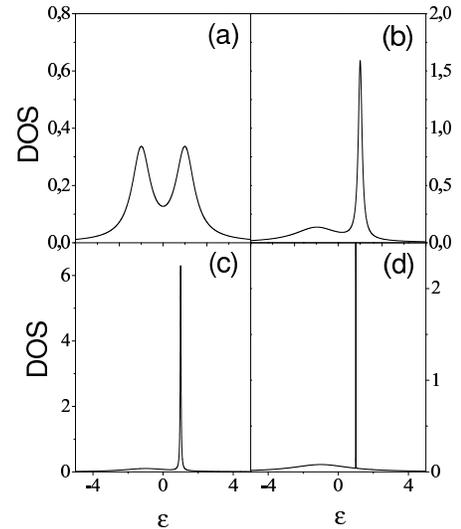}
\caption[Fig3]{Density of states $\rho $ as a function of the Fermi energy,
for $\Delta \varepsilon =0$, $t_{c}=\gamma _{1}$ and different values of $%
\gamma _{2}$: (a) $\gamma _{2}=0$, (b) $\gamma _{2}=0.3\gamma _{1}$, (c) $%
\gamma _{2}=0.6\gamma _{1}$ and (d) $\gamma _{2}=\gamma _{1}$.}
\label{Fig3}
\end{figure}

Additional insight into the physics underlying these results can be obtained
by examining the local density of states in each of the quantum dots. For $%
\Delta \varepsilon =0$ the density of states at both quantum dots
is the same, and the electron has the same probability of being
found at one or the other (covalent limit). It is straightforward
to show that $\rho _{1,2}=\rho $ is a superposition of two
Lorentzians of widths $\Gamma _{+}$ and $\Gamma _{-}$ at the
bonding and antibonding energies, respectively,
\begin{equation}
\rho (\varepsilon )=\frac{1}{2\pi }\left[ \frac{\Gamma _{+}}{%
(\varepsilon +t_{c})^{2}+\Gamma _{+}^{2}}+\frac{\Gamma _{-}}{(\varepsilon
-t_{c})^{2}+\Gamma _{-}^{2}}\right] .  \label{eq-21}
\end{equation}
Fig. \ref{Fig3} shows this quantity as a function of the energy
for the same parameters used in Fig. \ref{Fig2}. As we see, the
local density of states is the simple superposition of two
Lorentzians, one wide and one narrow as one approaches the
symmetric configuration. For $\gamma _{2}=\gamma _{1}$, the
Lorenztian centered at the bonding energy acquires the width
$2\gamma _{1}$, while the one at the antibonding energy becomes a
delta function. This last result indicates that the antibonding
state turns totally localized in the molecule. In fact, it is
straightforward to show that the Green function of this state ($
G_{aa}^{r}(t)=-i\theta (t)\langle \{d_{a}(t),d_{a}^{\dagger
}(0)\}\rangle$, where $d_{a}=(d_{1}-d_{2})/\sqrt{2}$) obeys the
equation of motion of an isolated state.

\begin{figure}[h]
\centering
\includegraphics[scale=0.28,angle=0]{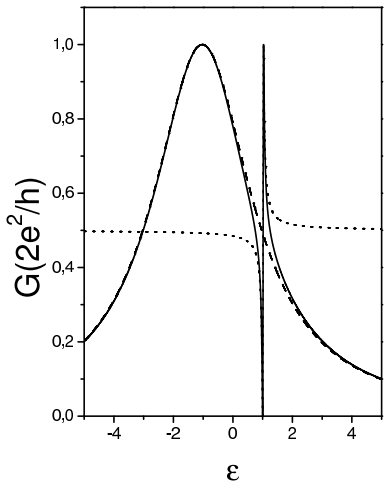}
\caption[Fig4]{Conductance as a function of the Fermi energy, for
$\Delta \varepsilon =\gamma _{1}/2$, $t_{c}=\gamma _{1}$ and
$\gamma _{2}=\gamma _{1} $. The dash and dotted lines are the
Lorentzian and the Fano line shape of Eqs. (\ref{eq-25}) and
(\ref{eq-25}), respectively. } \label{Fig4}
\end{figure}

\begin{figure}[ht]
\centering
\includegraphics[scale=0.28,angle=0]{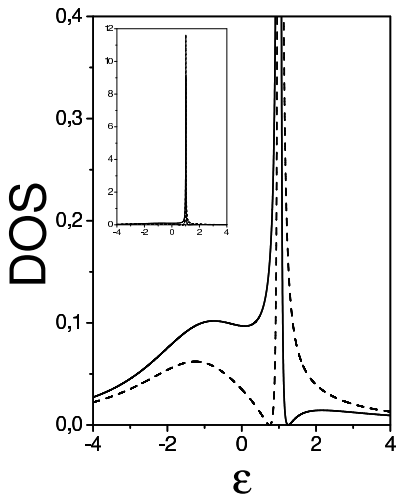}
\caption[Fig5]{ Local densities of states $\rho _{1}$ (solid line)
and $\rho _{2}$ (dash line) as a function of the Fermi energy, for
$\Delta \varepsilon =\gamma _{1}/2$, $t_{c}=\gamma _{1}$ and
$\gamma _{2}=\gamma _{1} $. } \label{Fig5}
\end{figure}

Now, let us discuss how a finite difference in the energies of the
different quantum dots modifies the results found in the fully
symmetrical situation. When  $\varepsilon_1 \neq \varepsilon_2$
and $\gamma_{2}=\gamma _{1}$ the conductance takes the form
\begin{equation}
G(\varepsilon )=\frac{2e^{2}}{h}\frac{4\gamma _{1}^{2}(t_{c}-\varepsilon
)^{2}}{C_{2}}  \label{eq-22}
\end{equation}
with
\begin{equation}
C_{2}=\left[ (\frac{\Delta \varepsilon }{2})^{2}+(t_{c}-\varepsilon
)(t_{c}+\varepsilon )\right] ^{2}+4\gamma _{1}^{2}(t_{c}-\varepsilon )^{2}.
\label{eq-23}
\end{equation}
Now the antiresonance takes place in $\varepsilon _{A}=t_{c}$ and
the resonances are further apart than for $\Delta \varepsilon =0$,
as one would expect, with their positions depending quadratically
on $\Delta \varepsilon$. As before, the conductance is a
convolution of a Lorentzian at the bonding energy and a Fano line
shape at the antibonding energy. For $\Delta \varepsilon \lesssim
t_{c}$ the Lorentzian has a constant line broadening $2\gamma
_{1}$,
\begin{equation}
G(\varepsilon )\simeq \frac{2e^{2}}{h}\frac{4\gamma _{1}^{2}}{(\varepsilon
-\varepsilon _{b})^{2}+4\gamma _{1}^{2}},  \label{eq-24}
\end{equation}
with $\varepsilon _{b}=-[t_{c}+(\Delta \varepsilon
)^{2}/(8t_{c})]$. The Fano line shape has a width dependent of the
difference of energies of the quantum dots levels, $\Gamma
_{a}=\gamma _{1}(\Delta \varepsilon )^{2}/[8(t_{c}^{2}+\gamma
_{1}^{2})]$,
\begin{equation}
G(\varepsilon )\simeq \frac{2e^{2}}{h}\frac{\gamma _{1}^{2}}{\gamma
_{1}^{2}+t_{c}^{2}}\frac{(e_{a}+Q)^{2}}{1+e_{a}^{2}},  \label{eq-25}
\end{equation}
where $e_{a}=[\varepsilon -t_{c}(1+\Gamma _{a}/\gamma
_{1})]/\Gamma _{a}$ and $Q=t_{c}/\gamma _{1}$. Fig. \ref{Fig4}
shows the conductance for $\Delta \varepsilon =\gamma _{1}/2$ and
$t_{c}=\gamma _{1}$, together with the corresponding pure
Breit-Wigner and Fano curves given by Eqs. (\ref{eq-24}) and
(\ref{eq-25}). Notice that the line broadening of the Fano peak
depends quadratically on the value of $\Delta \varepsilon$, but
differences of energy $\Delta \varepsilon$ of the order of
$\gamma_1$ result in Fano peaks narrow enough; for instance, if
$\Delta \varepsilon=\gamma_1=t_c$, $\Gamma_a=\gamma_1/16 \ll
\gamma_1$.
The local densities of states $\rho _{1}$ and $\rho _{2}$ are, respectively,
\begin{equation}
\rho _{1}(\varepsilon )=\frac{1}{\pi }\frac{(\Delta \varepsilon
-2t_{c}+2\varepsilon )^{2}}{4C_{2}},  \label{eq-26}
\end{equation}
\begin{equation}
\rho _{2}(\varepsilon )=\frac{1}{\pi }\frac{(\Delta \varepsilon
+2t_{c}-2\varepsilon )^{2}}{4C_{2}},  \label{eq-27}
\end{equation}
and, as observed in Fig.\ref{Fig5}, are the superposition of a
Breit-Wigner resonance close to the bonding energy and a Fano line
shape around the antibonding energy, similarly to the conductance.
It can be seen from the above analysis that for quantum dots with different
energies, both molecular states always contribute to the conductance.

\section{Conclusions}

In this work, we studied the conductance and density of states at
zero temperature of a quantum dot molecule attached to leads in
the range between a connection in series and a symmetric parallel
configuration. We found that the conductance is composed of
Breit-Wigner and Fano line shapes at the bonding and antibonding
energies, respectively, with their line broadenings controlled by
the asymmetry of the configuration. The narrowing (widening) of a
line in the conductance can be interpreted as an increase
(reduction) of the lifetime of the corresponding molecular state.
From the densities of states it can be deduced that the
antibonding state becomes progressively more localized as the
asymmetry of the configuration is reduced. Surprisingly, when the
configuration is completely symmetrical the tunneling through the
antibonding state is totally suppressed and the bonding state is
the only one participating in the transmission. In this limit, the
antibonding  becomes a coherent localized state with zero
localization length. The strong reduction of the decoherence
processes exhibited by the present system may have applications in
quantum computing.

\begin{acknowledgments}
F.\ C.\ and M. L. L. d. G. were supported in part by C\'atedra
Presidencial en Ciencia, and F.\ C.\ also received support from
FONDECYT, grant 1020829. P.\ A.\ O.\ would like to thank financial
support from Milenio ICM P99-135-F and FONDECYT under grant
1020269.
\end{acknowledgments}


\begin{thebibliography}{99}


\bibitem{qd1}  M.\ A.\ Kastner, Rev. Mod. Phys. \textbf{64}, 849 (1992).

\bibitem{qd2}
P. L. McEuen {\em et al.}, Phys. Rev. Lett. {\textbf 66}, 1926
(1991); R. C. Ashoori {\em et al.}, {\em ibid.} {\textbf 71}, 613
(1993); P. L. McEuen, Science {\textbf 276}, 1729 (1997).



\bibitem{ABoscill} A. Yacoby {\em et al.}, Phys. Rev. Lett. {\textbf 74}, 4047 (1995);
R. Schuster {\em et al.}, Nature {\textbf 385}, 417 (1997); E.
Buks {\em et al.}, {\em ibid} {\textbf 391}, 871 (1998); Yang Ji
{\em et al.} Science {\textbf 290}, 779 (2000).


\bibitem{kondo}  D.\ Goldhaber-Gordon {\em et al.}, Nature (London) \textbf{391}%
, 156 (1998); D.\ Goldhaber-Gordon {\em et al.}, Phys.\ Rev.\
Lett.\ \textbf{81}, 5225 (1998); S.\ M.\ Cronenwett {\em et al.},
Science \textbf{281}, 540 (1998); W. G. van der Wiel {\em et al.},
{\em ibid} {\textbf 289}, 2105 (2000).



\bibitem{fano} J. G\"{o}res {\em et al.}, Phys. Rev. B {\textbf 62}, 2188 (2000);
A. A. Clerk, X. Waintal, P. W. Brouwer, Phys. Rev. Lett. {\textbf
86}, 4636 (2001); I. G. Zacharia, D. Goldhaber-Gordon, G. Granger,
M. A. Kastner, Yu. B. Khavin, Hadas Shtrikman, D. Mahalu and U.
Meirav, Phys. Rev. B {\textbf 64}, 155311 (2001); Kensuke
Kobayashi, Hisashi Aikawa, Shingo Katsumoto and Yasuhiro Iye,
Phys. Rev. Lett. {\textbf 88}, 256806 (2002).



\bibitem{blick2}  R. H. Blick, D. Pfannkuche, R. J. Haug, K. v. Klitzing and
K. Eberl, Phys. Rev. Lett. \textbf{80} 4032 (1998).

\bibitem{jeong}  H. Jeong, A. M. Chang and M. R. Melloch, Science
\textbf{293} 2221 (2001).

\bibitem{holleitner1} A. W. Holleitner, C. R. Decker, H. Qin, K.
Eberl and R. H. Blick, Phys. Rev. Lett. {\textbf 87}, 256802
(2001).

\bibitem{holleitner2} A. W. Holleitner, R. H. Blick, A. K. H\"{u}ttel,
K. Eberl and J. P. Kotthaus, Science {\textbf 297}, 70 (2002).


\bibitem{schedelbeck} G. Schedelbeck, W. Wegscheider, M. Bichler
and G. Abstreiter, Science {\textbf 278}, 1792 (1997).

\bibitem{oosterkamp}  T. H. Oosterkamp, T. Fujisawa, W. G. van der Wiel, K.
Ishibashi. R. V. Hijman, S. Tarucha and Leo. P. Kouwenhoven, Nature \textbf{%
395} 873 (1998).


\bibitem{tmols}  G. Klimeck, G. Chen and S. Datta, Phys. Rev. B
\textbf{50} 2316 (1994); Cheng Niu, Li-jun Liu and Tsung-han Lin,
{\em ibid.} {\textbf 51}, 5130 (1995); C. A. Stafford and S. D.
Darma, Phys. Rev. Lett. {\textbf 72}, 3590 (1994); A. Aharony, O.
Entin-Wohlman and Y. Imry, Phys. Rev. B {\textbf 61}, 5452 (2000);
A. Aharony, O. Entin-Wohlman, Y. Imry and Y. Levinson, Phys. Rev.
B {\textbf 62}, 13561 (2000); Ram\'on Aguado and David C.
Langreth, Phys. Rev. Lett. {\textbf 85}, 1946 (2000); Pedro A.
Orellana, G. A. Lara, Enrique V. Anda, Phys. Rev.B \textbf{65}
155317 (2002).

\bibitem{Tdqd1} Tae-Suk Kim and S. Hershfield, Phys. Rev. B \textbf{63}, 245326
(2001); Y. Takazawa, Y. Imai and N. Kawakami, Cond-mat/0205016v1.

\bibitem{kang}  Kicheon Kang and Sam Young Cho, Cond-mat/0210009v1.

\bibitem{boese} Daniel Boese, Walter Hofstetter, Herbet Schoeller, Phys.
Rev. B \textbf{66} 125315 (2002).

\bibitem{loss1} Daniel Loss and David DiVincenzo, Phys. Rev. A
{\textbf 57}, 120 (1998).

%
\bibitem{blick3} R. H. Blick, A. K. H\"{u}ttel, A. W. Holleitner,
E. M. H\"{o}hberger, H. Qin, J. Kirschbaum, J. Weber, W.
Wegscheider, M. Bichler, K. Eberl and J. P. Hotthaus, to appear in
Physica E.


\bibitem{datta} Supriyo Datta, ``Electronic transport in mesoscopic
systems'' (Cambridge Univ. Press, 1997).



\bibitem{ddotp1}  Bj\"{o}rn Kubala and J\"{u}rgen K\"{o}nig, Phys. Rev. B
\textbf{65}, 245301.

\bibitem{2channel}  T.\ V.\ Shahbazyan and M.\ E.\ Raikh, Phys. Rev. B
\textbf{49}, 17123 (1994).

\end{thebibliography}
\end{document}